\documentclass[11pt,a4paper]{article}
\usepackage{latexsym}
\usepackage{amssymb}
\usepackage{amsmath}
\usepackage{authblk}
\title { Warm inflation with an oscillatory inflaton in the non-minimal kinetic coupling model}
\author[1]{Parviz Goodarzi\thanks{goodarzi@ut.ac.ir}}
\author[2]{H. Mohseni Sadjadi \thanks{mohsenisad@ut.ac.ir}}
\affil[1]{Department of Science, University of Ayatollah Ozma Borujerdi }
\affil[2]{Department of Physics, University of Tehran}

\begin{document}
\maketitle

\begin{abstract}
In the cold inflation scenario, the slow roll inflation and reheating via coherent rapid oscillation, are usually considered as two distinct eras.
When the slow roll ends, a rapid oscillation phase begins and the inflaton decays to relativistic particles reheating the Universe.
In another model dubbed warm inflation, the rapid oscillation phase is suppressed, and we are left with only a slow roll period during which the reheating occurs.
Instead, in this paper, we propose a new picture for inflation in which the slow roll era is suppressed and only the rapid oscillation phase exists.
Radiation generation during this era is taken into account, so we have warm inflation with an oscillatory inflaton.
To provide enough e-folds, we employ the non-minimal derivative coupling model. We study the cosmological perturbations
and compute the temperature at the end of warm oscillatory inflation.
\end{abstract}

\section{Introduction}

In the standard inflation model, the accelerated expansion and the reheating epochs are two distinct eras
\cite{guth,inflaton1,Liddle}. But in the warm inflation, relativistic particles are produced during the slow roll.
Therefore, the warm inflation explains the slow roll and onset of the radiation dominated era in a unique framework \cite{Berera,Berera2}.
Warm inflation is a good model for large scale structure formation, in which the density fluctuations arise
from thermal fluctuation \cite{Berera3,Berera4}. Various models have been proposed for warm inflation, e.g. tachyon warm inflation,
warm inflation in loop quantum cosmology, etc. \cite{Herrera1,Herrera2,Zhang}.

Oscillating inflation was first introduced in \cite{Mukhanov}, where it was proposed that the inflation may continue, after the slow roll, during rapid coherent oscillation in the reheating era. An expression for the corresponding number
of e-folds was obtained in\cite{Liddle}.

Scalar field oscillation in inflationary model was also pointed out briefly in \cite{ref}, where the decay of scalar fields during their oscillations to inflaton particles was proposed.

A brief investigation of the adiabatic perturbation in the oscillatory inflation can be found in \cite{Taruya}.
The formalism used in \cite{Mukhanov} was extended in \cite{Lee}, by considering a coupling between inflaton and the Ricci scalar curvature. The shape of the potential, required to end the oscillatory inflation, was investigated in \cite{sami}. The rapid oscillatory phase provides a few e-folds so we cannot ignore the slow roll era in this formalism. Due to small few number of e-folds, a detailed study of the evolution of quantum fluctuations has not been performed.
To cure this problem, one can consider nonminimal derivative coupling model. The cosmological aspects of this model have been widely studied in the literature \cite{sushkov}.

The oscillatory inflation in the presence of a non-minimal kinetic coupling was studied in \cite{sadjadi2} and there was shown that in high-friction regime, the non-minimal coupling increases the e-folds number and so can remedy the problem of
the smallness of the number of e-folds arising in \cite{Mukhanov}. Scalar and tensor perturbations and power spectrum and spectral index for scalar and tensor modes in oscillatory inflation, were derived in \cite{sadjadi2}, in agreement with Planck 2013 data. However, it is not clear from this scenario how reheating occurs or the Universe becomes radiation dominated after the end of inflation. For non-minimal derivative coupling model, the reheating process after the slow roll and warm slow roll inflation are studied in \cite{good1, good2} and \cite{nozari1,nozari2} respectively.

In the present work, inspired by the models mentioned above,  we will consider oscillatory inflation in non-minimal derivative coupling model.

We will assume that the inflaton decays to the radiation during the oscillation,  providing a new scenario:  warm oscillatory inflation. Equivalently, this can be viewed as an oscillatory reheating phase which is not preceded by the slow roll.

In the second section, we examine conditions for warm oscillatory inflation and study the evolution of energy density of the scalar field and radiation.
In the third section, the thermal fluctuation is considered and spectral index and power spectrum are computed.
We will consider observational constraints on oscillatory warm inflation parameters by using Planck 2015 data \cite{planck}. In the fourth section, the temperature at the end of warm inflation is calculated. We will compute tensor perturbation in the fifth section and in the last section, we conclude our results.

We use units $\hbar=c=1$ throughout this paper.

\section{Oscillatory warm inflation}

In this section, based on our previous works \cite{sadjadi2,good1,good2}, we will introduce rapid oscillatory inflaton decaying to radiation in non-minimal kinetic coupling model. We start with the action \cite{Germani1}
\begin{equation}\label{1}
S=\int \Big({M_P^2\over 2}R-{1\over 2}\Delta^{\mu \nu}\partial_\mu
\varphi \partial_{\nu} \varphi- V(\varphi)\Big) \sqrt{-g}d^4x+S_{int}+S_{r},
\end{equation}
where $\Delta^{\mu \nu}=g^{\mu \nu}+{1\over M^2}G^{\mu \nu}$, $G^{\mu \nu}=R^{\mu \nu}-{1\over 2}Rg^{\mu \nu}$ is the Einstein
tensor, $M$ is a coupling constant with mass dimension, $M_P=2.4\times 10^{18}GeV$ is the reduced Planck
mass, $S_{r}$ is the radiation action and $S_{int}$ describes the interaction of the scalar field with radiation.
There are not terms containing more than two times derivative, so we have not additional degrees of freedom in
this theory. We can calculate energy momentum tensor by variation of action with respect to the metric
\begin{equation}\label{2}
T_{\mu\nu}=T^{(\varphi)}_{\mu\nu}+{1\over M^2}\Theta_{\mu\nu}+T^{(r)}_{\mu\nu}.
\end{equation}
The energy momentum tensor for radiation is
\begin{equation}\label{3}
T^{(r)}_{\mu\nu}=(\rho_r+P_r)u_{\mu}u_{\nu}+P_rg_{\mu\nu},
\end{equation}
where $u^{\mu}$ is the four-velocity of the radiation and $T^{(\varphi)}_{\mu\nu}$ is the minimal
coupling counterpart of the energy momentum tensor
\begin{equation}\label{4}
T^{(\varphi)}_{\mu\nu}=\nabla_{\mu}\varphi\nabla_{\nu}\varphi-{1\over2}g_{\mu\nu}{(\nabla\varphi)}^2-g_{\mu\nu}V(\varphi).
\end{equation}
The energy momentum tensor corresponding to the non-minimal coupling term is
\begin{eqnarray}\label{5}
&&\Theta_{\mu\nu}=-{1\over2}G_{\mu\nu}{(\nabla\varphi)}^2-{1\over2}R\nabla_{\mu}\varphi\nabla_{\nu}\varphi+
R^{\alpha}_{\mu}\nabla_{\alpha}\varphi\nabla_{\nu}\varphi\\ \nonumber
&&+R^{\alpha}_{\nu}\nabla_{\alpha}\varphi\nabla_{\mu}\varphi+R_{\mu\alpha\nu\beta}\nabla^{\alpha}\varphi\nabla^{\beta}\varphi
+\nabla_{\mu}\nabla^{\alpha}\varphi\nabla_{\nu}\nabla_{\alpha}\varphi\\\nonumber
&&-\nabla_{\mu}\nabla^{\nu}\varphi\Box \varphi
-{1\over2}g_{\mu\nu}\nabla^{\alpha}\nabla^{\beta}\varphi\nabla_{\alpha}\nabla_{\beta}\varphi+{1\over2}g_{\mu\nu}{(\Box\varphi)}^2\\\nonumber
&&-g_{\mu\nu}\nabla_{\alpha}\varphi\nabla_{\beta}\varphi R^{\alpha\beta}.
\end{eqnarray}
Energy transfer between the scalar field and radiation is assumed to be
\begin{equation}\label{6}
Q_{\mu}=-\Gamma u^{\nu} \partial_{\mu}\varphi \partial_{\nu} \varphi,
\end{equation}
where
\begin{equation}\label{7}
\nabla^{\mu}T^{(r)}_{\mu\nu}=Q^{\nu} \qquad and \qquad \nabla^{\mu}(T^{(\varphi)}_{\mu\nu}+{1\over M^2}\Theta_{\mu\nu})=-Q_{\nu}.
\end{equation}

The scalar field equation of motion, in Friedmann-Lema\^{\i}tre-Robertson-Walker (FLRW) metric is
\begin{equation}\label{8}
(1+{3H^2\over M^2})\ddot{\varphi}+3H(1+{3H^2\over M^2}+{2\dot{H}\over M^2})\dot{\varphi}+V'(\varphi)+\Gamma\dot{\varphi}=0,
\end{equation}
where $H={\dot{a}\over a}$ is the Hubble parameter, a "dot" is differentiation with respect to cosmic time $t$, prime denotes differentiation with respect to the scalar field $\varphi$.

$\Gamma$ is a positive constant, first introduced in \cite{gamma} as a phenomenological term which describes the decay of $\varphi$ to the radiation during reheating era. This term was vastly used in the subsequent literature studying the inflaton decay in the reheating era (see \cite{gamma1} and \cite{kolb} and the references therein), where like our model the inflaton experiences a rapid oscillation phase. In \cite{gamma2}, it is shown that the production of particles during high-frequency regime in reheating era can be expressed by adding a polarization term to the inflaton mass. To do so,
a Lagrangian comprising the inflaton field and its interactions with bosonic and fermionic fields was employed. It was shown that the phenomenological term proposed in \cite{gamma1} can be derived in this context. The precise form of the dissipative term depends on the coupling between inflaton and relativistic particles it decays to, and also on the interactions of relativistic particles. As the nature of inflaton and these relativistic particles are not yet completely known, the precise form  of $\Gamma$  is not clear.

However one may employ a phenomenological effective field theory or also thermal field theory \cite{gamma3}, to study the effective dependency of $\Gamma$ on temperature and dynamical fields. Thermal effect can also be inserted  by including the thermal correction in the equations of motion \cite{gamma3}. In the period where the inflaton is dominant over relativistic thermal particles, it is safe to approximately take $\Gamma$  as $\Gamma=\Gamma\Big|_{T=0}$ (like \cite{gamma}), as explained in \cite{gamma4}.

Similarly, in the framework of slow roll warm inflation, the possibility that $\Gamma$ is a function of $\varphi$ and temperature, was discussed in the literature\cite{Berera2,Xiao}.

The Friedmann equations are given by
\begin{eqnarray}\label{9}
H^2&=&{1\over 3M_P^2}(\rho_\varphi+\rho_{r}) \nonumber \\
 \dot{H}&=&-{1\over 2M_P^2}(\rho_\varphi+\rho_{r}+P_\varphi+P_{r}).
\end{eqnarray}

The energy density and the pressure of inflaton can be expressed as
\begin{equation}\label{10}
\rho_\varphi=(1+{9H^2\over M^2}){\dot{\varphi}^2\over 2}+V(\varphi),
\end{equation}
and
\begin{equation}\label{11}
P_\varphi=(1-{3H^2\over M^2}-{2\dot{H}\over M^2}){\dot{\varphi}^2\over 2}-V(\varphi)-{2H\dot{\varphi}\ddot{\varphi}\over M^2},
\end{equation}
respectively. Energy density of radiation is $\rho_{r}={3\over4}TS$ \cite{Berera2}. $S$ is the entropy density and $T$ is the temperature.
The equation of state parameter for radiation is ${1\over3}$, hence the rate of radiation production is given by
\begin{equation}\label{12}
\dot{\rho_{r}}+4H\rho_{r}=\Gamma{\dot{\varphi}}^2.
\end{equation}

We assume that the potential is even, $ V(-\varphi)=V(\varphi)$, and consider rapid oscillating solution (around $\varphi=0$) to (\ref{8}), which in high-friction regime, $H^2\gg M^2$,  reduces to
\begin{equation}\label{r1}
\ddot{\varphi}+3H\left(1+{2\over 3}{\dot{H}\over H^2}\right)\dot{\varphi}+{M^2V_{,\varphi}\over 3H^2}+{M^2\Gamma\over 3H^2}=0.
\end{equation}
In our formalism the inflation has a quasi periodic evolution
\begin{equation}\label{r2}
\varphi(t)=\phi(t) \cos(\int A(t) dt),
\end{equation}
with time dependent amplitude $\phi(t)$. The rapid oscillation (or high frequency oscillation) is characterized by
\begin{equation}\label{r3}
\left|{\dot{H}\over H}\right|\ll A,\,\,\,\left|{\dot{\phi}\over \phi}\right|\ll A,\,\,\,\,\left|{\dot{\rho_{\varphi}}\over \rho_{\varphi}}\right|\ll A.
\end{equation}
The existence of such a solution is verified in \cite{good1}. It is worth to note that for a power law potential $V(\varphi)=\lambda \varphi^q$, (\ref{r3}) holds provided that
\begin{equation}\label{r4}
\Phi\ll \left(q^2M_P^4 M^2\over \lambda \right)^{1\over q+2},
\end{equation}
which is opposite to the slow roll condition $\varphi^{q+2}\gg \left(M_P^4M^2\over \lambda\right)$ \cite{good1}.

The period of oscillation is
\begin{equation}\label{13}
\tau(t)=2\int_{-\phi}^\phi {d\varphi(t)\over \dot{\varphi}(t)},
\end{equation}
and the rapid oscillation occurs for $H\ll {1\over \tau}$ and ${\dot{H}\over H}\ll {1\over \tau}$.
The inflaton energy density may estimated as  $\rho_{\varphi}=V(\phi(t))$. In this epoch $\rho_\varphi$ and $H$ change insignificantly during a period of oscillation in the sense indicated in (\ref{r3}).

In rapid oscillatory phase, the time average of adiabatic index, defined by $\gamma={\rho_\varphi + P_\varphi\over \rho_\varphi}$ is given by $\gamma=\big<{\rho_\varphi+P_\varphi\over \rho_\varphi}\big>$, where bracket denotes time averaging over one oscillation
\begin{equation}\label{r5}
\big<O(t)\big>={\int_t^{t+\tau}O(t')dt'\over \tau}.
\end{equation}
For a power law potential
\begin{equation}\label{14}
V(\varphi)=\lambda\varphi^q,
\end{equation}
and in high-friction limit $({H^2\over M^2}\gg1)$, the adiabatic index becomes \cite{sadjadi2}
\begin{equation}\label{15}
\gamma\approx{2q\over 3q+6}.
\end{equation}

By averaging the continuity equation, we obtain \cite{good1}
\begin{equation}\label{16}
<\rho_\varphi\dot{>}+3H\gamma<\rho_\varphi>+{\gamma\Gamma M^2 \over 3H^2}<\rho_\varphi>=0.
\end{equation}

when the Universe is dominated by $\varphi$-particles, we take
\begin{equation}\label{R1}
\Gamma\ll {9H^3\over M^2}.
\end{equation}
By this assumption the radiation may be still in equilibrium, and besides we can neglect the third term in (\ref{16}). But as $H$ decreases, and the radiation production term becomes more relevant, this approximation fails and the third terms in (\ref{16}) get the same order of magnitude as the second term at a time $t_{rh}$.  Note that at $t_{rh}$ the radiation and inflaton densities have the same order of magnitude $\rho_{\varphi}(t_{rh})\sim \rho_{r}(t_{rh})$ \cite{good1}, \cite{kolb}. When $t< t_{rh}$, the average of energy density of the scalar field can be approximated as
\begin{equation}\label{17}
\big<\rho_\varphi\big>\propto a(t)^{-3\gamma}.
\end{equation}

By using relation (\ref{17}) and the Friedmann equation ($H^2\approx{1\over 3M_P^2}\rho_\varphi$), in the $\varphi$ dominated era, we can easily obtain
\begin{equation}\label{18}
a(t)\propto t^{{q+2\over q}}\propto t^{{2\over 3\gamma}}.
\end{equation}

Therefore the Hubble parameter in the inflaton dominated era can be estimated as $H\approx {2\over3\gamma t}$ . In the rapid oscillation phase and with the power law potential (\ref{14}) we can write the amplitude of the oscillation as
\begin{equation}\label{19}
\phi(t)\propto a(t)^{-{2\over q+2}}\propto t^{-{2\over q}}.
\end{equation}

Our formalism is similar to methods used in the papers studying the reheating era after inflation in the minimal case \cite{kolb}. But in the minimal case, for $\Gamma<<3H$ until $\Gamma\sim H$, where the Universe is dominated by the oscillating inflaton, instead of (\ref{18}), we have $a(t)\propto t^{2\over 3}$.

In high-friction limit, time averaging over one oscillation gives
\begin{equation}\label{r6}
\big<\dot{\varphi}^2(t)\big>={2M^2\over 9H^2(t)}\big<\rho_\varphi(t)-V(\varphi(t))\big>,
\end{equation}
where we have used that the Hubble parameter changes insignificantly during one period of oscillation. But
\begin{eqnarray}\label{r7}
\big<\rho_\varphi(t)-V(\varphi(t))\big>&=&{\int_{-\phi(t)}^{\phi(t)}\sqrt{\rho_\varphi-V(\varphi)}d\varphi\over \int_{-\phi(t)}^{\phi(t)} {d\varphi \over \sqrt{\rho_\varphi-V(\varphi)}}}\nonumber \\
&=&\lambda \phi^q(t) {\int_{0}^{1}\sqrt{1-x^q}dx\over \int_{0}^{1} {dx \over \sqrt{1-x^q}}}\nonumber \\
&=&\lambda \phi^q(t){q\over q+2},
\end{eqnarray}
therefore
\begin{equation}\label{20}
\big<\dot{\varphi}^2\big>\approx\gamma M_P^2 M^2.
\end{equation}
This relation shows that for non-minimal derivative coupling model and in the rapid oscillation phase, when the Universe is $\varphi$ dominated, $\big<\dot{\varphi}^2\big>$ is approximately a constant. By inserting (\ref{20}) into the equation (\ref{12}) we obtain
\begin{equation}\label{20.1}
\rho_r={3\Gamma \gamma^2 M^2 M_P^2\over (8+3\gamma)}t\bigg[1-({t_{0}\over t})^{(1+{8\over3\gamma})}\bigg],
\end{equation}
where $t_{0}$ is the time at which $\rho_r=0$. The number of e-folds from a specific time $t_{*}\in (t_0,t_{RD})$ in inflation until radiation dominated epoch, is given by
\begin{equation}\label{20.2}
\mathcal{N}_I=\int_{t_{*}}^{t_{RD}}Hdt\approx\int_{t_{*}}^{t_{RD}}{2\over3\gamma t}dt\approx{2\over3\gamma}\ln\bigg({t_{RD}\over t_{*}}\bigg),
\end{equation}
where $t_{RD}$ is the time at which the universe becomes radiation dominated and inflation ceases. At this time
\begin{equation}\label{20.4}
\rho_r(T_{RD})\approx\rho_{\varphi}(t_{RD}).
\end{equation}
We can calculate the temperature at the end of warm inflation by \cite{Berera}
\begin{equation}\label{20.3}
\rho_r(t_{RD})=g_{RD}{\pi^2\over30}T_{RD}^4,
\end{equation}
where $g_{RD}$ is number of degree of freedom of relativistic particles and $T_{RD}$ is the temperature of radiation at the beginning of radiation dominated era.

\section{Cosmological perturbations}

In this section, we study the evolution of thermal fluctuation during oscillatory warm inflation.   We use the framework used in \cite{Berera3} and ignore the possible viscosity terms and
shear viscous stress \cite{visineli}.
To investigate cosmological perturbations, we split the metric into two components: the background and the perturbations. The background is described by homogeneous and isotropic FLRW metric with oscillatory scalar field and the perturbed sector of the metric determines anisotropy. We assume that the radiation is in thermal equilibrium during warm inflation.
The thermal fluctuations arising in warm inflation evolve gradually via cosmological perturbations equations. Until the freeze out time, the thermal noise has not a significant effect                                                                                                                                                 on perturbations development \cite{Berera3}.
We consider the evolution equation of the first order cosmological perturbations for a system containing inflaton and radiation. In the longitudinal gauge the metric can be written as \cite{Weinberg}.
\begin{equation}\label{21}
ds^2=-(1+2\Phi)dt^2+a^2(1-2\Psi)\delta_{ij}dx^idx^j.
\end{equation}
As mentioned before, the energy momentum tensor splits into radiation part $T^{\mu\nu}_r$ and inflaton part $T^{\mu\nu}_{\varphi}$ as
\begin{equation}\label{22}
T^{\mu\nu}=T^{\mu\nu}_r+T^{\mu\nu}_{\varphi}.
\end{equation}
The unperturbed parts of four velocity components of the radiation fluid satisfy $\overline{u}_{ri}=0$ and $\overline{u}_{r0}=-1$. By using normalization condition $g^{\mu\nu}u_{\mu}u_{\nu}=-1$, the perturbed part of the time component of the four velocity becomes
\begin{equation}\label{24}
\delta u^0=\delta u_0={h_{00}\over 2}.
\end{equation}
The space components $\delta u^i$,  are independent dynamical variables and $\delta u_i=\partial_i\delta u$ \cite{Weinberg}.
Energy transfer is described by \cite{Moss}
\begin{equation}\label{25}
 Q_{\mu}=-\Gamma u^{\nu}\partial_{\mu}\varphi \partial_{\nu}\varphi.
\end{equation}
We have also
 \begin{equation}\label{26}
\nabla_{\mu}T^{\mu\nu}_r=Q^{\nu},
\end{equation}
and
\begin{equation}\label{27}
\nabla_{\mu}T^{\mu\nu}_{\varphi}=-Q^{\nu}.
\end{equation}
(\ref{25}) gives $Q_{0}=\Gamma \dot{\varphi}^2$ and the unperturbed equation (\ref{26}) becomes
 $Q_{0}=\dot\rho_r +3H(\rho_r+P_r)$  which is the continuity equation for the radiation field.
 In the same way Eq.(\ref{27}) becomes $-Q_{0}=\dot\rho_{\varphi} +3H(\rho_{\varphi}+P_{\varphi})$.
Perturbations to the energy momentum transfer are described by  (there is no perturbation for the dissipation factor $\Gamma$ which we have assumed to be a constant)
\begin{equation}\label{28}
\delta Q_{0}=-\delta\Gamma\dot{\varphi}^2+\Phi\Gamma\dot{\varphi}^2-2\Gamma\dot{\varphi}\dot{\delta\varphi}
\end{equation}
and
\begin{equation}\label{29}
\delta Q_{i}=-\Gamma\dot{\varphi}\partial_i{\delta\varphi}.
\end{equation}

The variation of the equation (\ref{26}) is $\delta(\nabla_{\mu}T^{\mu\nu}_r)=\delta Q^{\nu}$, so its (0-0) component is
\begin{equation}\label{30}
\dot{\delta\rho_r}+4H\delta\rho_r+{4\over3}\rho_r\nabla^2\delta u-4\dot{\Psi}\rho_r=
-\Phi\Gamma{\dot\varphi}^2+\delta\Gamma{\dot\varphi}^2+2\Gamma\dot{\delta\varphi}\dot{\varphi}.
\end{equation}
Similarly, for the $i-th$ component we derive
\begin{equation}\label{31}
4\rho_r\dot{\delta u^i}+4\dot{\rho_r}\delta u^i+20 H\rho_r\delta
u^i=-[3\Gamma\dot{\varphi}\partial_i\delta\varphi+\partial_i\delta\rho_r+4\rho_r\partial_i\Phi].
\end{equation}

The equation of motion for $\delta\varphi$, computed by variation of (\ref{27}), is  $\delta(\nabla_{\mu}T^{\mu\nu}_{\varphi})=-\delta Q^{\nu}$. The zero component of this equation is
\begin{eqnarray}\label{32}
(1+{3H^2\over M^2})\ddot{\delta\varphi}+[(1+{3H^2\over M^2}+{2\dot{H}\over M^2})3H+\Gamma]\dot{\delta\varphi}
+\delta V'(\varphi)+\dot{\varphi}\delta\Gamma\\\nonumber
-(1+{3H^2\over M^2}+{2\dot{H}\over M^2}){\nabla^2\delta\varphi\over a^2}=\\\nonumber
-[2V'(\varphi)+3\Gamma\dot{\varphi}-{6H\dot{\varphi}\over M^2}(3H^2+2\dot{H})
-{6H^2\ddot{\varphi}\over M^2}]\Phi\\\nonumber
+(1+{9H^2\over M^2})\dot{\varphi}\dot\Phi+{2H\dot{\varphi}\over M^2}{\nabla^2\Phi\over a^2}\\\nonumber
+3(1+{9H^2\over M^2}+{2\dot{H}\over M^2}+{2H\ddot{\varphi}\over M^2})\dot{\Psi}+{6H\dot{\varphi}\over M^2}\ddot{\Psi}-{2(\ddot{\varphi}+H\dot{\varphi})\over M^2}{\nabla^2\Psi\over a^2}.
\end{eqnarray}

The $0-0$ component of the perturbation of the Einstein equation $G_{\mu\nu}=-8\pi GT_{\mu\nu}$ is
\begin{eqnarray}\label{33}
-3H\dot{\Psi}-3H^2\Phi+{\nabla^2\Psi\over a^2}=4\pi G\big[-(1+{18H^2\over M^2}){\dot{\varphi}}^2\Phi-{9H{\dot{\varphi}}^2\over M^2}\dot{\Psi} \\\nonumber
+{{\dot{\varphi}}^2\over M^2}{\nabla^2\Psi\over a^2}+\acute{V(\varphi)}\delta\varphi+(1+{9H^2\over M^2})\dot\varphi\dot{\delta\varphi}
-{2H\dot{\varphi}\over M^2}{\nabla^2{(\delta\varphi)}\over a^2}+\delta\rho_r\big],
\end{eqnarray}
and its $i-i$ component is
\begin{eqnarray}\label{34}
&&(3H^2+2\dot{H})\Phi+H(3\dot{\Psi}+\dot{\Phi})+{\nabla^2(\Phi-\Psi)\over 3a^2}+\ddot{\Psi}=\\\nonumber
&&4\pi G[({(3H^2+2\dot{H}){2\dot{\varphi}^2\over M^2}-{\dot{\varphi}}^2+{8H\dot{\varphi}\ddot{\varphi}\over M^2}})\Phi+
{3H{\dot{\varphi}}^2\over M^2}\dot{\Phi} \\\nonumber
&&+{{\dot{\varphi}}^2\over M^2}{\nabla^2\Phi\over 3a^2}+({3H{\dot{\varphi}}^2\over M^2}+{2\dot{\varphi}\ddot{\varphi}\over M^2})\dot{\Psi}+{{\dot{\varphi}}^2\over M^2}\ddot{\Psi}+{{\dot{\varphi}}^2\over M^2}{\nabla^2\Psi\over 3a^2} \\\nonumber
&&-\acute{V(\varphi)}\delta\varphi-[(-1+{3H^2\over M^2}+{2\dot{H}\over M^2})\dot{\varphi}+{2H\ddot\varphi\over M^2}]\dot{\delta\varphi}\\\nonumber
&&-{2H\dot{\varphi}\over M^2}\ddot{\delta\varphi}
+{2(\ddot{\varphi}+H\dot{\varphi})\over M^2}{\nabla^2{(\delta\varphi)}\over 3a^2}+\delta P_r].
\end{eqnarray}

By using $-H\partial_i\Phi-\partial_i\dot{\Psi}=4\pi G(\rho+P)\partial_i\delta u$,  we can obtain (from $(0-i)$ component of the field equation)
\begin{eqnarray}\label{35}
&&H\Phi+\dot{\Psi}=4\pi G[{3H\dot{\varphi}^2\over M^2}\Phi+{\dot{\varphi}^2\over M^2}\dot{\Psi}+(1+{3H^2\over M^2})\dot{\varphi}\delta\varphi-{2H\dot{\varphi}\over M^2}\dot{\delta\varphi}\\\nonumber
&&+(\rho_r+P_r)\delta u].
\end{eqnarray}
 Using(\ref{30}-\ref{35}) we can calculate perturbation parameters.

 Depending on the physical process, e.g. thermal noise, expansion, curvature fluctuations, three separate regimes for the evolution of the scalar field fluctuations may be
 considered \cite{Berera3}. But one can generalize this approach, by adding stochastic noise source and viscous terms to cosmological perturbations equations \cite{visineli}.

 During inflation the background has two components, oscillatory scalar field and radiation.
The energy density of the scalar field decreases due to expansion and radiation generation.
Quantities related to the scalar field in the background have oscillatory behaviors. So we replace the background quantities with their average values over oscillation.  Also, we consider non-minimal derivative coupling at the high-friction limit.

 By going to the Fourier space, the spatial parts of perturbational quantities get $e^{ikx}$ where $k$ is the wave number. So $\partial_j\rightarrow ik_j $ and $\nabla^2\rightarrow-k^2 $. Also we define
\begin{equation}\label{37}
\delta u=-{a\over k}v e^{ikx}.
\end{equation}
So (\ref{30}) becomes
\begin{equation}\label{38}
\dot{\delta\rho_r}+4H\delta\rho_r+{4\over3} ka\rho_r v-4\rho_r\dot{\Phi}=-\Gamma M^2 M_P^2\Phi,
\end{equation}
and (\ref{31}) becomes
\begin{equation}\label{39}
4{a\over k}(\dot{(\rho_r v)}+4H(\rho_r v))=-\delta\rho_r-4\rho_r\Phi.
\end{equation}
(\ref{32}) reduces to
\begin{eqnarray}\label{40}
&&({3H^2\over M^2})\ddot{\delta\varphi}+[({3H^2\over M^2}+{2\dot{H}\over M^2})3H+\Gamma]\dot{\delta\varphi}
+\delta V'(\varphi)=\\\nonumber
&&-2V'(\varphi)\Phi+3({9H^2\over M^2}+{2\dot{H}\over M^2})\dot{\Phi}.
\end{eqnarray}
From  (\ref{33}) we have
\begin{equation}\label{41}
-3H\dot{\Phi}(1-{3\gamma\over2})-3H^2\Phi(1-3\gamma)={1\over{2M_P^2}}(V'(\varphi)\delta\varphi+\delta\rho_r),
\end{equation}
and rewrite  (\ref{34}) as
\begin{equation}\label{42}
(3H^2+2\dot{H})\Phi(1-\gamma)+H\dot{\Phi}(4-3\gamma)+\ddot{\Phi}(1-{1\over2}\gamma)={1\over{2M_P^2}}(-V'(\varphi)\delta\varphi+\delta P_r).
\end{equation}

Note that we have replaced $\dot{\phi}^2$ and $\dot{\phi}$ by their average values i.e. $<\dot{\phi}^2>=\gamma M^2 M_P^2$ and $<\dot{\phi}>=0$. We restrict ourselves to the high-friction regime ${H^2\over M^2}\gg 1$ and the modes satisfying
${k\over a}\ll H$ and the zero-shear gauge $\Phi=\Psi$ \cite{Berera3} are considered.

(\ref{35}) may be written as
\begin{equation}\label{43}
H\Phi(1-{3\over2}\gamma)+\dot{\Phi}(1-{1\over2}\gamma)=-{2\over{3M_P^2}}{a\over k}(v\rho_r),
\end{equation}
and the time derivative of (\ref{35}) gives
\begin{equation}\label{44}
(H\dot{\Phi}+\dot{H}\Phi)(1-{3\over2}\gamma)+\ddot{\Phi}(1-{1\over2}\gamma)=-{2\over{3M_P^2}}{a\over k}(H(v\rho_r)+\dot{(v\rho_r)}).
\end{equation}

By analyzing the above equations we find
\begin{equation}\label{45}
[3H^2(1-{3\over2}\gamma-{1\over3}\gamma)+\dot{H}({2\over3}-{7\over6}\gamma)]\Phi+{5\over6}(4-3\gamma)H\dot{\Phi}+{5\over6}(1-{1\over2}\gamma)\ddot{\Phi}=0.
\end{equation}

During the rapid oscillation, the Hubble parameter is $H={2\over 3\gamma t}$,  therefore (\ref{45}) becomes
\begin{equation}\label{46}
({2\over3\gamma})[{2\over\gamma}-{13\over3}+{7\over6}\gamma]{\Phi\over t^2}+{5\over9\gamma}(4-3\gamma){\dot{\Phi}\over t} +{5\over6}(1-{1\over2}\gamma)\ddot{\Phi}=0.
\end{equation}
This equation has the solution $\Phi\propto t^{\alpha_\pm}$, therefore
\begin{equation}\label{47}
({2\over3\gamma})[{2\over\gamma}-{13\over3}+{7\over6}\gamma]+{5\over9\gamma}(4-3\gamma)\alpha +{5\over6}(1-{1\over2}\gamma)\alpha(\alpha-1)=0.
\end{equation}
$\alpha's$ are the roots of this quadratic equation. We denote the positive root by $\alpha_+$. From equations (\ref{41}) and (\ref{42}), we deduce
\begin{equation}\label{48}
-{1\over{M_P^2}}V'(\varphi)\delta\varphi=2(3H^2(1-2\gamma)+\dot{H}(1-\gamma))\Phi+(7-{3\over 2}\gamma)H\dot{\Phi}+(1-{1\over2}\gamma)\ddot{\Phi}.
\end{equation}
It is now possible to use relation $\Phi\propto t^{\alpha_+}$  to obtain $\delta\varphi$
\begin{eqnarray}\label{49}
&&-{1\over{M_P^2}}V'(\varphi)\delta\varphi= \nonumber \\ &&\bigg[{4\over3\gamma}({2\over\gamma}-5+\gamma)+{2\over 3\gamma}(7-{15\gamma\over2})\alpha_{+}
+(1-{1\over2}\gamma)\alpha_{+}(\alpha_{+}-1)\bigg]{\Phi\over t^2}.
\end{eqnarray}
$\delta\varphi$ simplifies to
\begin{eqnarray}\label{50}
&&\delta\varphi=-C{{M_P^{2+\alpha_{+}}}\over V'(\varphi)}t^{\alpha_{+}-2}\times \nonumber \\ &&\bigg[{4\over3\gamma}({2\over\gamma}-5+\gamma)+{2\over 3\gamma}(7-{15\gamma\over2})\alpha_{+}
+(1-{1\over2}\gamma)\alpha_{+}(\alpha_{+}-1)\bigg]
\end{eqnarray}
where $C$ is a numerical constant. Thus the density perturbation, from relation (\ref{50}), becomes \cite{dp}
\begin{eqnarray}\label{51}
&&\delta_H\approx {16\pi\over 5 M_P^{2+\alpha_{+}}} \nonumber \\
&&{V'\delta\varphi\over\bigg[{4\over3\gamma}({2\over\gamma}-5+\gamma)+{2\over 3\gamma}(7-{15\gamma\over2})\alpha_{+}
+(1-{1\over2}\gamma)\alpha_{+}(\alpha_{+}-1)\bigg]t^{\alpha_{+}-2}}.
\end{eqnarray}

In this relation $\delta\varphi$ is the scalar field fluctuation during the warm inflation, which instead of quantum fluctuation,  are generated by thermal fluctuation \cite{Berera,gamma3}.   Due to the thermal fluctuations, $\varphi$ satisfies the Langevin equation with a stochastic noise source, using which one finds
\cite{nozari1}
\begin{equation}\label{52}
\delta\varphi^2={k_FT\over2\pi^2},
\end{equation}
where $k_F$ is the freeze out scale, containing also terms corresponding to the non-minimal coupling. To compute $k_F$, we must determine when the damping rate of relation (\ref{42})
becomes less than the expansion rate $H$. At $t_F$ (freeze out time \cite{Berera3}), the freeze out wave number $k_F={k\over a(t_F)}$ is given by \cite{nozari1}
\begin{equation}\label{53}
k_F=\sqrt{\Gamma H+3H^2(1+{3H^2\over M^2})}.
\end{equation}
In the minimal case ${H^2\over M^2}=0$, and (\ref{53}) gives the well known result \cite{gamma3}.
\begin{equation} \label{ref123}
\delta\varphi^2={\sqrt{\Gamma H+3H^2}T\over 2\pi^2},
\end{equation}
which  reduces to $\delta\varphi^2={\sqrt{3}H T\over 2\pi^2}$ \cite{Berera} in weak dissipative regime $\Gamma\ll H$, and to $\delta\varphi^2={\sqrt{\Gamma H}T\over 2\pi^2}$, in the strong dissipative regime $\Gamma\gg H$. For a more detailed discussion about the scalar field fluctuations(\ref{ref123}), based on quantum field theory first principles, see \cite{ref}.
In our case, as we are restricted to the the high-friction regime ${H^2\over M^2}\gg 1$ and also use the approximation (\ref{R1}) before the radiation dominated era, we have
\begin{equation}\label{R2}
\delta\varphi^2={3H^2T\over2 M\pi^2}.
\end{equation}
Note that our study is restricted to  the region $H<\Gamma \lesssim\left({H^2\over 9M^2}\right)H$.
By using (\ref{R2}), the density perturbation
\begin{eqnarray}\label{54}
&&\delta_H^2\approx {\left({16\pi\over 5M_P^{2+\alpha_+}}\right)}^2 t^{4-2\alpha_{+}}\times \nonumber \\
&&{V'^2\delta\varphi^2\over\bigg[{4\over3\gamma}({2\over\gamma}-5+\gamma)+{2\over 3\gamma}(7-{15\gamma\over2})\alpha_{+}
+(1-{1\over2}\gamma)\alpha_{+}(\alpha_{+}-1)\bigg]^2}
\end{eqnarray}
can be rewritten as
\begin{eqnarray}\label{55}
&&\delta_H^2\approx {\left({128\over 25M_P^{4+2\alpha_+}}\right)}t^{4-2\alpha_{+}}\left({3H^2\over M}\right)T\times
\nonumber \\
&&{V'^2\over\bigg[{4\over3\gamma}({2\over\gamma}-5+\gamma)+{2\over 3\gamma}(7-{15\gamma\over2})\alpha_{+}
+(1-{1\over2}\gamma)\alpha_{+}(\alpha_{+}-1)\bigg]^2}.
\end{eqnarray}
We can now calculate power spectrum from relation $P_s(k_0)={25\over4}\delta_H^2(k_0)$ \cite{dp}. $k_0$ is a pivot scale.
The spectral index for scalar perturbation is
 \begin{equation}\label{56}
n_s-1={d\ln{\delta_H^2}\over d\ln{k}}.
\end{equation}
The derivative is taken at horizon crossing $k\approx aH$.
The spectral index may be written as
 \begin{equation}\label{57}
n_s-1={d\ln{\delta_H^2}\over d\ln{(aH)}}=\bigg({1\over H+{\dot{H}\over H}}\bigg){d\ln{\delta_H^2}\over dt}.
\end{equation}
From $H={2\over3\gamma t}$ we have
 \begin{equation}\label{58}
n_s-1\approx \bigg({t\over{2\over 3\gamma}-1}\bigg){d\ln{\delta_H^2}\over dt},
\end{equation}
therefore
\begin{equation}\label{59}
n_s-1\approx \bigg({4\over3\gamma}-{5\over2}-2\alpha_{+}\bigg)\bigg({1\over{2\over 3\gamma}-1}\bigg).
\end{equation}

This relation gives the spectral index as a function of $\gamma$. From Planck 2015 data $n_s=0.9645\pm0.0049$
(68\% CL, Planck TT,TE,EE+lowP) $\gamma$ is determined as $\gamma=0.55902\pm0.00016$.

\section{Evolution of the Universe and temperature of the warm inflation}

In this section, by using our previous results,  we intend to calculate the temperature of warm inflation as a function of observational parameters for the power law potential (\ref{14}) and a constant dissipation coefficient $\Gamma $, in high-friction limit.
For this purpose we follow the steps introduced in  \cite{Mielczarek1}, and divide the evolution of the Universe from $t_*$ (a time at which a pivot scale exited the Hubble radius) in inflation era until now into three parts \\
$I-$ from $t_\star$ until the end of oscillatory warm inflation, denoted by $t_{RD}$. in this period energy density of the oscillatory scalar field is dominated.\\
$II-$ from $t_{RD}$ until recombination era, denoted by $t_{rec}$.\\
$III-$ from $t_{rec}$ until the present time $t_0$.\\
Therefore the number of e-folds from horizon crossing until now becomes
\begin{eqnarray}\label{60}
\mathcal{N}&=&\ln{({a_0\over a_\star})}=\ln{({a_0\over a_{rec}})}+\ln{({a_{rec}\over a_{RD}})}+\ln{({a_{RD}\over a_{\star}})}=\\\nonumber
&&\mathcal{N}_{I}+\mathcal{N}_{II}+\mathcal{N}_{III}
\end{eqnarray}
\subsection{Oscillatory warm inflation}

During the warm oscillatory inflation, the scalar field oscillates and decays into the ultra-relativistic particles. In this period the energy density of oscillatory scalar field is dominated and the Universe expansion is accelerated.
The beginning time of radiation dominated era is determined by the condition $\rho_r(t_{RD})\simeq\rho_{\varphi}(t_{RD})$ which gives \cite{good1,good2}
\begin{equation}\label{62}
{t_{RD}}^3={4(8+3\gamma)\over9\Gamma\gamma^4M^2}.
\end{equation}
From equations (\ref{62}) and (\ref{20.1}) we can calculate energy density of radiation at $t_{RD}$
\begin{equation}\label{63}
\rho_r(t=t_{RD})\approx M_P^2\bigg[{12\Gamma^2\gamma^2M^4\over{(8+3\gamma)}^2} \bigg]^{1\over3}.
\end{equation}
 Note that $t_{RD}\sim t_{rh}$, where $t_{rh}$ is defined after (\ref{R1}). The temperature of the Universe at the end of oscillatory warm inflation becomes
\begin{equation}\label{64}
{T_{RD}}^4\approx {30M_P^2\over\pi^2g_{RD}}\bigg[{12\Gamma^2\gamma^2M^4\over{(8+3\gamma)}^2} \bigg]^{1\over3}.
\end{equation}

\subsection{Radiation dominated and recombination eras}
At the end of the warm inflation the magnitude of radiation energy density equals the energy density of the scalar field. Thereafter the universe enters a radiation dominated era. During this period, the Universe is filled with ultra-relativistic particles which are in thermal equilibrium. In this epoch the Universe undergoes an adiabatic expansion where the entropy per comoving volume is conserved: $dS=0$ \cite{kolb}. In this era the entropy
density, $s=Sa^{-3}$, is \cite{kolb}
\begin{equation}\label{65}
s={2\pi^2\over 45}g T^3.
\end{equation}
So we have
\begin{equation}\label{66}
{a_{rec}\over a_{RD}}={T_{RD}\over T_{rec}}\left({g_{RD}\over g_{rec}}\right)^{1\over 3}.
\end{equation}
In the recombination era, $g_{rec}$ corresponds to degrees of freedom of photons, hence $g_{rec}=2$. Thus
\begin{equation}\label{67}
\mathcal{N}_{II}= \ln\left({T_{RD}\over
T_{rec}}\left({g_{RD}\over 2}\right)^{1\over 3}\right).
\end{equation}

By the expansion of the Universe, the temperature decreases via
$T(z)=T(z=0)(1+z)$, where $z$ is the redshift parameter. Hence $T_{rec}$ in terms of $T_{CMB}$  is
\begin{equation}\label{68}
T_{rec}=(1+z_{rec})T_{CMB}.
\end{equation}

We have also
\begin{equation}\label{69}
{a_0\over a_{rec}}=(1+z_{rec}).
\end{equation}
Therefore
\begin{equation}\label{70}
\mathcal{N}_{II}+\mathcal{N}_{III}=\ln\left({T_{RD}\over
T_{CMB}}\left({g_{RD}\over 2}\right)^{1\over3}\right).
\end{equation}

\subsection{Temperature of the warm oscillatory inflation}

To obtain temperature of the warm inflation we must
determine  $\mathcal{N}$ in (\ref{60}). We take $a_0=1$, so the number of
e-folds from the horizon crossing until the present time is
$\Delta=\exp(\mathcal{N})$, where
\begin{equation}\label{71}
\Delta={1\over a_*}={H_*\over k_0}\approx {2\over 3\gamma t_* k_0}.
\end{equation}
By relations (\ref{71},\ref{70},\ref{60}) we can derive $T_{RD}$,
\begin{equation}\label{72}
T_{RD}=T_{CMB}{({2\over g_{RD}})}^{1\over3}{2\over3\gamma k_0}\bigg[{4(8+3\gamma)\over9\Gamma\gamma^4M^2} \bigg]^{-{2\over9\gamma}}
\times t_{*}^{({2\over3\gamma}-1)}.
\end{equation}
We can remove $\Gamma M^2$ in this relation by (\ref{64})
\begin{equation}\label{72.1}
T_{RD}^{(1-{4\over3\gamma})}\approx {2T_{CMB}\over3\gamma k_0}{({2\over g_{RD}})}^{{\gamma-1\over3\gamma}}\bigg[{2\sqrt{5}M_P\over\pi\gamma} \bigg]^{-{2\over3\gamma}}\times t_{*}^{({2\over3\gamma}-1)}.
\end{equation}
By using relation $\mathcal{P}_s(k_0)={25\over4}\delta_H^2(k_0)$ and equation (\ref{55}), power spectrum becomes
\begin{eqnarray}\label{73}
&&\mathcal{P}_s(k_0)\approx \bigg({32\over M_P^{4+2\alpha_{+}}}\bigg) \times \nonumber \\
&&{\left<V'(\varphi_*)\right>^2\over\bigg[{4\over3\gamma}({2\over\gamma}-5+\gamma)+{2\over 3\gamma}(7-{15\gamma\over2})\alpha_{+}
+(1-{1\over2}\gamma)\alpha_{+}(\alpha_{+}-1)\bigg]^2}\nonumber \\
&&\times t_*^{4-2\alpha_{+}}{\sqrt{\Gamma H_*+3H_*^2(1+{3H_*^2\over M^2})}}T_*.
\end{eqnarray}
In this relation $T_\star$ is the temperature of the universe at the horizon crossing. By relation (\ref{20.1}) we can calculate temperature at horizon crossing as a function of $t_*$
\begin{equation}\label{74}
T_{*}=\bigg[ {90\Gamma\gamma^2M^2M_P^2\over(8+3\gamma)\pi^2g_*}\bigg]^{1\over4}t_*^{1\over4}.
\end{equation}
We can remove $\Gamma M^2$ in relation (\ref{74}) by (\ref{64})
\begin{equation}\label{75}
T_{*}=\bigg[ {\pi\gamma g_{RD}^{1\over2}\over2\sqrt{10}M_P}\bigg]^{1\over4} T_{RD}^{3\over2} t_*^{1\over4}.
\end{equation}
We have taken $g_{RD}\sim g_{*}$. The time average of the potential derivative may computed as follows
\begin{eqnarray}\label{76}
<V'>&=&q\lambda{\int_{-\phi}^{\phi}\varphi ^{q-1}{d\varphi \over \dot{\varphi}}\over \int_{-\phi}^{\phi}{d\varphi \over \dot{\varphi}}}\nonumber \\
&=&q\lambda \phi^{n-1}{\int_0^1 {x^{q-1}dx\over \sqrt{1-x^q}}\over \int_0^1 {dx\over \sqrt{1-x^q}}}\nonumber \\
&=&2q\lambda{\Gamma\left({2+q\over 2q}\right)\over \Gamma \left({1\over q}\right)}\phi^{q-1}.
\end{eqnarray}
In inflationary regime we have $H^2\approx{1\over 3M_P^2}\rho_\varphi\approx {1\over 3M_P^2}\lambda \phi^q$ and $H^2\approx {4\over 9 \gamma^2 t^2}$, therefore
\begin{equation}\label{76.1}
<V'(\varphi_*)>=12\lambda{\gamma\over2-3\gamma}{\Gamma\left({1\over 3\gamma}\right)\over \Gamma \left({1\over 3\gamma}-{1\over2}\right)}\left({4M_P^2\over 3\lambda \gamma^2}\right)^{9\gamma-2\over 6\gamma}t_*^{2-9\gamma\over 3\gamma}.
\end{equation}
Thus we can write (\ref{73}) as
\begin{equation}\label{77}
\mathcal{P}_s(k_0)\approx M_P^{({5\over2}-{4\over3\gamma}-2\alpha_{+})}\lambda^{({2\over3\gamma}-1)}\Gamma^{1\over4}M^{1\over2}g_{RD}^{-{1\over4}}\beta t_*^{(-{11\over4}+{4\over3\gamma}-2\alpha_{+})}.
\end{equation}
$\beta$ is given by
\begin{equation}\label{78}
\beta={{2048\over \sqrt{\pi}}\left({90\gamma^2\over 8+3\gamma}\right)^{1\over 4}{\left({4\over 3\gamma^2}\right)}^{9\gamma-2\over3\gamma}\bigg({\Gamma\left({1\over 3\gamma}\right)\over \Gamma \left({1\over3\gamma}
-{1\over 2}\right)}\bigg)^2\over(2-3\gamma)^2\bigg[{4\over3\gamma}({2\over\gamma}-5+\gamma)+{2\over 3\gamma}(7-{15\gamma\over2})\alpha_{+}
+(1-{1\over2}\gamma)\alpha_{+}(\alpha_{+}-1)\bigg]^2}.
\end{equation}
From equation (\ref{77}), we derive $t_*$ as
\begin{equation}\label{78.1}
t_*=\bigg[{\mathcal{P}_s(k_0)g_{RD}^{1\over4}\over M_P^{({5\over2}-{4\over3\gamma}-2\alpha_{+})}\lambda^{({2\over3\gamma}-1)}M^{1\over 2}\Gamma^{1\over4}\beta}\bigg]^{12\gamma\over16-33\gamma-24\gamma\alpha_+}.
\end{equation}

By substituting $t_*$ from relation (\ref{78.1}) into equation (\ref{72}), the temperature at the end of warm oscillatory inflation
or beginning of the radiation domination is obtained as
\begin{eqnarray}\label{79}
T_{RD}=T_{CMB}{({2\over g_{RD}})}^{1\over3}{2\over3\gamma k_0}\bigg[{4(8+3\gamma)\over9\Gamma\gamma^4M^2} \bigg]^{-{2\over9\gamma}}
\times \\\nonumber
\bigg[{\mathcal{P}_s(k_0)g_{RD}^{1\over4}\over M_P^{({5\over2}-{4\over3\gamma}-2\alpha_{+})}\lambda^{({2\over3\gamma}-1)}M^{1\over 2}\Gamma^{1\over4}\beta}\bigg]^{4(2-3\gamma)\over16-33\gamma-24\gamma\alpha_+}.
\end{eqnarray}
The number of e-folds during warm oscillatory inflation becomes
\begin{equation}\label{80}
\mathcal{N}_I\approx{2\over3\gamma}\ln \left(\left({4(8+3\gamma)\over9\Gamma\gamma^4M^2}\right)^{1\over 3}
\left({\mathcal{P}_s(k_0)g_{RD}^{1\over4}\over M_P^{({5\over2}-{4\over3\gamma})-2\alpha_{+}}\lambda^{({2\over3\gamma}-1)}M^{1\over 2}\Gamma^{1\over4}\beta}\right)^{{12\gamma\over-16+33\gamma+24\gamma\alpha_+}}\right).
\end{equation}

We set $g_{RD}=106.75$, which is the ultra relativistic degrees of freedom at the electroweak energy scale.
Also, from Planck 2015 data, at the pivot scale $k_0=0.002Mpc^{-1}$ and in one sigma level, we have
 $\mathcal{P}_s(k_0)=(2.014\pm0.046)\times10^{-9}$ and $n_s=0.9645\pm0.0049$ $(68\% CL,Planck TT,TE,EE+lowP)$ \cite{planck}.
By using $\gamma=0.55902 $, $M=10^{-16}M_P$, $\lambda=(10^{-8}M_P)^{4-q}$,  and $\Gamma=10^{-4}M_P$ in equation (\ref{79})
the temperature of the universe at the end of warm inflation and the number of e-folds become
$T_{end}\approx3.83\times10^{12}GeV$, and $\mathcal{N}=61.42$ respectively.

 \section{Tensorial perturbation}

In this part, we follow the method used in \cite{Kaushik} to study tensorial perturbation. The power spectrum for tensorial perturbation
is given by \cite{Kaushik}
\begin{equation}\label{81}
{P_{t}(k)}={k^3\over 2\pi^2}|{v_{k}\over z}|^2\coth({k\over2T}),
\end{equation}
where $v_k$ can be calculated from Mukhanov equation \cite{Germani2}
\begin{equation}\label{82}
{d^2v_k\over d\eta^2}+\big(c^2k^2-{1\over z}{d^2z\over d\eta^2}\big)v_k=0.
\end{equation}
$\eta$ is the conformal time, $c_t$ is the sound speed for tensor mode and $k$ is wave number for mode function $v_k$ \cite{Germani2} and $z$ is given by
\begin{equation}\label{83}
z=a(t)M_p{\sqrt{e_{ij}^\lambda e_{ij}^\lambda }\over2}\sqrt{1-\alpha}.
\end{equation}
The tensor of polarization is normalized as $e_{ij}^\lambda e_{ij}^{\lambda^\prime}=2\delta_{\lambda\lambda^\prime} $. For our model, with a quasi periodic scalar inflaton background, we have
$\alpha={\dot{\varphi}^2\over 2M^2M_p^2}$ and $c$ is given by relation
\begin{equation}\label{84}
c^2={1+\alpha\over 1-\alpha}.
\end{equation}
Therefore
\begin{equation}\label{85}
{1\over z}{d^2z\over d\eta^2}=({q\over2}+1)({q\over2}+2)\eta^{-2}.
\end{equation}
By using this relation, the equation for mode function becomes
\begin{equation}\label{86}
{d^2v_k\over d\eta^2}+\big(c^2k^2-({q\over2}+1)({q\over2}+2)\eta^{-2}\big)v_k=0.
\end{equation}
Solution to this mode equation are the Hankel functions of the first and second kind
\begin{equation}\label{87}
v_k(\eta)=\eta^{1\over 2}[C^{(1)}(k)H_\nu^{(1)}(ck\eta)+C^{(2)}(k)H_\nu^{(2)}(ck\eta)].
\end{equation}
Well within the horizon, the modes satisfy $k\gg aH$, and can be approximated by flat waves. Therefore
\begin{equation}\label{88}
v_{k}(\eta)\approx{\sqrt{\pi}\over 2}e^{i(\nu+{1\over 2}){\pi\over 2}}{ (-\eta)}^{1\over 2}H_\nu^{(1)}(-ck\eta).
\end{equation}

On the other hand, when we want to compute power spectrum, we need to have modes that are outside the horizon. So by taking the limit ${k\over aH}\rightarrow0$, we obtain the asymptotic form of mode function as
\begin{equation}\label{89}
v_{k}(\eta)\rightarrow e^{i(\nu+{1\over 2}){\pi \over2 } }2^{(\nu-{3\over2})}{\Gamma(\nu)\over\Gamma({3\over2})}{1\over\sqrt{2ck}}{(-ck\eta)}^{(-\nu+{1\over2})}.
\end{equation}
By using  this relation we can write the power spectrum as
\begin{equation}\label{90}
P_t(k)={k^3\over 2\pi^2}{2^{(2\nu-3)}\over\beta^2 a^2}\bigg({\Gamma(\nu)\over\Gamma({3\over2})}\bigg)^2{1\over2c k}
{(-ck\eta)}^{(-2\nu+1)}\coth({k\over2T}).
\end{equation}
In the rapid oscillation epoch $\epsilon={\dot{H}\over H^2}={3\gamma\over 2}$  (see (\ref{18})), so we can write the conformal time as
\begin{equation}\label{91}
\eta=-{1\over aH}{1\over 1-\epsilon}.
\end{equation}
At the horizon crossing $c_sk=aH$,  we can write (\ref{90}) as
\begin{equation}\label{92}
P_{t}=A^2_{t}(q)\bigg({H\over M_p}\bigg)^2\coth({k\over2T})|_{ck=aH},
\end{equation}
where
\begin{equation}\label{93}
A_t(q)={3^{({1\over2})}{2^{(q-{1\over2})}}{\Gamma({3\over2}+{q\over2})}(q+2)^{(-{q+1\over2})}\over\pi\Gamma({3\over2})(q+3)^{({1\over4})}(2q+3)^{({3\over4})}}.
\end{equation}
The ratio of tensor to scalar spectrum (from relations (\ref{77}) and (\ref{92})) becomes
\begin{equation}\label{94}
r={P_t(k_0)\over \mathcal{P}_s(k_0)}\approx \bigg[{4A^2_{t}(q)g_{RD}^{1\over4}t_*^{({4\over3\gamma}-{11\over4}-2\alpha_{+})}\over {9\gamma^2 M_p^{({9\over2}-{4\over3\gamma}-2\alpha_{+})}\lambda^{({2\over3\gamma}-1)}M^{1\over2}\Gamma^{1\over4}}\beta }\bigg]\coth({k\over2T})\bigg|_{ck=aH}.
\end{equation}

From relation (\ref{94}) by using $\gamma=0.55902 $, $ g_{RD}=106.75$, $M=10^{-16}M_p$ and $\Gamma=10^{-4}M_p$,
the ratio of tensor to scaler at the pivot scale $k_*=0.002 Mpc^{-1}$ becomes $r\approx0.081$ which is consistent with planck 2015 data $r_{0.002}<0.10$ $(95\% CL,
Planck TT, TE, EE + lowP)$.

 \section{Conclusion}

In the standard model of inflation, inflaton begins a coherent rapid oscillation after the slow roll. During this stage, inflaton decays to radiation and reheats the Universe. In this paper, we considered a rapid oscillatory inflaton during inflation era. This scenario does not work in minimal coupling model due to the fewness of e-folds during rapid oscillation. But the non-minimal derivative coupling can remedy this problem in high-friction regime. Therefore we proposed a new model in which inflation and rapid oscillation are unified without considering the slow roll. The number of e-folds was calculated. We investigated cosmological perturbations and the temperature of the Universe was determined as a function of the spectral index.

We used a phenomenological approach to describe the interaction term between the inflaton and the radiation, but a precise study about thermal radiation productions must be
derived from quantum field theory principles.  An attempt in this subject may be found in \cite{ref}.

To complete our study  one may consider quantum and thermal corrections to the parameters of the system such as the inflaton  mass and its coupling to the radiation. In the slow roll model, the role of theses corrections on the observed spectrum is studied in the literature \cite{quantum}. Recently in \cite{be}, it was shown that in the warm slow roll inflation, it is possible to sustain the flatness of the potential against the thermal and loop quantum corrections. In the non-minimal derivative coupling model, by power counting analysis, and unitarity constraint which implies $H\ll \Lambda$, where $\Lambda=(H^2M_P)^{1\over 3}$ is the cutoff of the theory, it was shown that for power-law potentials quantum radiation corrections are subleading \cite{Germani1}. However, it may be interesting to study in details the effect of loop quantum corrections and corresponding renormalization on the behavior of our model and on its spectral index and power spectrum.

 Note that our model is an initial study of warm oscillatory inflation. Further studies may be performed by considering thermal correction to the effective potential \cite{potential}, and also by taking into account the temperature dependency of the dissipative factor and checking all consistency conditions. We leave these problems as an outlook for future works.

\end{document}